\documentclass[submission, LectureNotes]{SciPost}

\usepackage{amsmath}
\usepackage{graphicx,multirow,subcaption}
\numberwithin{equation}{section}

\newcommand{\dr}[1]{\widetilde{#1}}

\newcommand{\e}{\text{e}}
\newcommand{\tr}{\text{tr}}
\renewcommand{\vec}[1]{\boldsymbol #1}
\newcommand{\im}{\text{i}}

\def\12{\frac{1}{2}}

\begin{document}

\begin{center}{\Large \textbf{
Transport in one-dimensional integrable quantum systems
}}\end{center}

\begin{center}
J. Sirker*
\end{center}

\begin{center}
Department of Physics and Astronomy, University of Manitoba, Winnipeg R3T 2N2, Canada
\\
* sirker@physics.umanitoba.ca
\end{center}

\begin{center}
\today
\end{center}


\section*{Abstract}
{\bf These notes are based on a series of three lectures given at the
Les Houches summer school on 'Integrability in Atomic and Condensed
Matter Physics' in August 2018. They provide an introduction into the
unusual transport properties of integrable models in the linear
response regime focussing, in particular, on the spin-$1/2$ XXZ spin
chain.}

\vspace{10pt}
\noindent\rule{\textwidth}{1pt}
\tableofcontents\thispagestyle{fancy}
\noindent\rule{\textwidth}{1pt}
\vspace{10pt}

\section{Outline}
In these lecture notes I will discuss transport in one-dimensional
quantum systems at finite temperatures in the linear response
regime. After a general introduction, a particular focus will be on
the unusual transport properties of integrable systems. Note that
these notes are not meant to be an exhaustive review of the research
field. They are based on the content of three lectures given at the
Les Houches summer school on 'Integrability in Atomic and Condensed
Matter Physics' in August 2018 and are therefore necessarily limited
in scope. I also note that these lectures on transport build on
material which was presented at the summer school in earlier
lectures. Foundations of the coordinate, algebraic, and thermodynamic
Bethe ansatz, in particular, are assumed to be known already.

In order to be concrete, I will mostly concentrate on a particular
integrable lattice model, the XXZ spin chain
\begin{equation}
\label{Ham}
H=\sum_\ell \left[J\left(S^x_\ell S^x_{\ell+1} + S^y_\ell S^y_{\ell+1} +\Delta S^z_\ell S^z_{\ell+1}\right)-h S^z_\ell \right] \, .
\end{equation}
Here $J$ is the exchange constant, $\Delta$ the exchange anisotropy,
and $h$ an external magnetic field. $S^\alpha$ are spin-1/2 operators
fulfilling the commutation relations
$[S^\alpha,S^\beta]=i\varepsilon_{\alpha\beta\gamma}S^\gamma$. In the
following, we will often parametrize the anisotropy as
$\Delta=\cos(\gamma)$.
It is also often useful to think about this model as a chain of
interacting spinless fermions
\begin{equation}
\label{Ham2}
H= \sum_\ell \left\{J\left[-\frac{1}{2}(c^\dagger_\ell c_{\ell+1} + h.c.) +\Delta \left(n_\ell-\frac{1}{2}\right)\left(n_{\ell+1}-\frac{1}{2}\right)\right]-h\left(n_\ell-\frac{1}{2}\right)\right\}
\end{equation}
with $n_\ell = c^\dagger_\ell c_\ell$ where $c_\ell$ is a fermionic
annihilation operator at site $\ell$. This alternative representation
is obtained by using the Jordan-Wigner transformation
\begin{equation}
\label{JW}
S^z_\ell \to n_\ell -\frac{1}{2},\quad S^+_\ell \to (-1)^\ell c^\dagger_\ell \e^{i\pi\phi_\ell},\quad S^-_\ell \to (-1)^\ell c_\ell \e^{-i\pi\phi_\ell} 
\end{equation}
with the ladder operators $S^{\pm}_\ell = S^x_\ell \pm \im S^y_\ell$ and
the Jordan-Wigner string $\phi_\ell=\sum_{j=1}^{\ell-1} n_j$. Note
that the Jordan-Wigner string does not show up explicitly in the
Hamiltonian (\ref{Ham2}) because the hopping is limited to
nearest-neighbour sites in the lattice.

In general, transport is a {\it non-equilibrium problem}: Spin
transport, for example, requires a magnetic field gradient while heat
transport is driven by a temperature gradient. In the following I
will, however, exclusively analyze the linear response regime using
{\it Kubo formulas}. In this regime, transport coefficients can be
obtained from dynamical correlation functions {\it calculated at
equilibrium}. The plan for the three lectures is then as follows: In
the first lecture, I will briefly recapitulate how currents and
transport coefficients can be defined. Furthermore, I will derive the
Mazur inequality and explain why integrability can lead to ballistic
transport even at finite temperatures. In the second lecture, the Kubo
formulas for the conductivities will be discussed. I will show, in
particular, how the Drude weight and the diffusion constant can be
obtained from real-time equilibrium current-current correlation
functions. Explicit results for the thermal Drude weight of the XXZ
chain will be derived. To obtain a broader physical understanding of
the interplay of ballistic and diffusive transport channels, I will
describe the XXZ chain at low energies using bosonization in the third
lecture. Finally, I will use the field theoretical description to
calculate the spin conductivity, obtain a concrete formula for the
spin diffusion constant, and discuss the general picture which emerges
from these calculations.

\section{Transport coefficients and linear response}
One way to derive the spin and thermal current operators for the XXZ
chain is based on a discrete version of the continuity equation where
the time derivative is calculated using the equation of motion. For
the total spin current $\mathcal{J}^s=\sum_\ell j^s_\ell$ we have, in
particular,
\begin{equation}
\label{cont}
\partial_t S^z_\ell = -\mbox{i}[S^z_\ell,H] = -(j^s_\ell-j^s_{\ell-1}) 
\end{equation}
leading to a current density
\begin{equation}
\label{spincurrent}
j^s_\ell = J(S^x_\ell S^y_{\ell+1} - S^y_\ell S^x_{\ell+1}) = \frac{\im J}{2}(S^+_\ell S^-_{l+1} - S^-_\ell S^+_{l+1}) \, .
\end{equation}
Using the Jordan-Wigner transformation (\ref{JW}) we see that in terms
of spinless fermions this corresponds to a particle current, i.e., the
difference between particles moving to the left and to the right.

Similarly, we can derive the thermal current operator
$\mathcal{J}^{\textrm{th}}=\sum_\ell j^{\textrm{th}}_\ell$ by the continuity equation
\begin{equation}
\label{cont1}
\partial_t h_{\ell,\ell+1} = -\mbox{i}[h_{\ell,\ell+1},H] = -(j^{\textrm{th}}_\ell-j^{\textrm{th}}_{\ell-1}) 
\end{equation}
where $H=H^0-h\sum_\ell S^z_\ell=\sum_\ell h_{\ell,\ell+1}=\sum_\ell
(h^0_{\ell,\ell+1}-hS^z_\ell)$. The thermal current thus splits into
two parts, $J^{\textrm{th}}=J^E-hJ^s$, where $J^s$ is the spin current
(\ref{spincurrent}) and $J^E$ the energy current obtained from the
continuity equation (\ref{cont1}) for the case of zero magnetic
field. In other words, at finite magnetic fields there is a
contribution to the thermal current due to particle
transport. Calculating the commutator in (\ref{cont1}) for $h=0$, leads
to an energy current density $j^E_\ell$ acting on three neighbouring
sites which can be written in compact form as
\begin{equation}
\label{ecurrent}
j^E_\ell = J^2 \sum_\ell \vec{S}_\ell \cdot (\vec{S}'_{\ell-1}\times \vec{S}'_{\ell+1}), \quad \vec{S}'_\ell = (S^x_\ell,S^y_\ell,\Delta S^z_\ell) \, .
\end{equation}

Alternatively, the spin current can also be derived by putting a flux
$\Phi$ through an XXZ ring in the fermionic formulation
(\ref{Ham2}). The flux then couples via the Peierls substitution
$c_\ell^\dagger c_{\ell+1} \to c_\ell^\dagger c_{\ell+1} \e^{-i
A_{\ell,\ell+1}}$. Here $A_{\ell,\ell+1}$ is the vector potential
along the bond with $\sum_\ell A_{\ell,\ell+1} =\Phi$. The current
operator is then given by $j^s_\ell = -\frac{\partial H}{\partial
A_{\ell,\ell+1}}\big|_{A\to 0}$. Furthermore, the diamagnetic term can
be obtained as $\frac{\partial^2 H}{\partial A^2}\big|_{A\to 0} =
H_{\textrm{kin}}$ where $H_{\textrm{kin}}$ is the hopping part of the
Hamiltonian (\ref{Ham2}).

The transport coefficients relate the currents to the gradients in temperature and magnetic field 
\begin{equation}
\label{Tcoeffs}
\left(\begin{array}{c} \mathcal{J}^{\textrm{th}} \\ \mathcal{J}^s \end{array}\right) = \left(\begin{array}{cc} \kappa_{\textrm{th}} & C^{\textrm{th}}_s \\ C^s_{\textrm{th}} & \sigma_s \end{array} \right)\left(\begin{array}{c} -\nabla T \\ \nabla h \end{array}\right)
\end{equation}
with $\kappa_{\textrm{th}}$ being the thermal conductivity and
$\sigma_s$ the spin conductivity. The coefficients $
C^{\textrm{th}}_s$ and $ C^s_{\textrm{th}}$ describe the creation of a
thermal current due to a magnetic field gradient and of a spin current
due to a thermal gradient, respectively. The latter is the spin
Seebeck effect which has been studied in much detail for ferromagnets
in the field of spintronics. From the Onsager relation
\cite{Mahan} it follows that $C^{\textrm{th}}_s = T C^s_{\textrm{th}}$. 

The, in general, complex and frequency dependent transport
coefficients are decomposed as, for example,
\begin{equation}
\label{Drude}
\sigma'_s(k=0,\omega) =2\pi D_s \delta(\omega) +\sigma^{\textrm{reg}}_s(\omega)
\end{equation}
where $\sigma'_s(k,\omega)$ denotes the real part of the spin
conductivity at momentum $k$ and frequency $\omega$. $D_s$ is the {\it
spin Drude weight}, and $\sigma^{\textrm{reg}}_s(\omega)$ the regular
part of the conductivity. We can write down a similar decomposition
for the thermal conductivity $\kappa_{\textrm{th}}(\omega)$. A
non-zero Drude weight signals {\it ballistic transport}, i.e.~a
diverging dc conductivity. Physically, this means that the current
does not completely relax. In a lattice system without impurities, we
expect this to happen at zero temperature where scattering processes
such as spin-spin or spin-phonon are frozen out. At finite
temperatures, on the other hand, we expect that in a generic clean
system the dc conductivity becomes finite. Scattering processes are
expected to lead to a temperature-dependent broadening of the delta
peak. This can only be avoided if a part of the current is fully
protected from relaxing by some conservation law.

This is the point where integrability comes into play. What makes
integrable models special, is that they have an {\it infinite set of
local conserved charges} $\mathcal{Q}_j$. Here we mean local in the
strict sense that
\begin{equation}
\label{local}
\mathcal{Q}_j=\sum_{\ell} q^j_\ell
\end{equation}
where $q^j$ is a local charge density acting on $j$ neighbouring
sites. For the XXZ chain, in particular, we can derive these charges
by defining a family of commuting transfer matrices
$[T(\theta),T(\theta')]=0$ with spectral parameter $\theta$. The local
conserved charges are then obtained by
\begin{equation}
\label{Charges}
\mathcal{Q}_{j+1} =\frac{d^{j}}{d\theta^{j}} \ln T(\theta)\big|_{\theta=1},\quad j\geq 1 \, .
\end{equation}
We refer to reference \cite{PereiraPasquier} for details. For now it
is just important to note that $\mathcal{Q}_2\propto H$ and
$\mathcal{Q}_3\propto
\mathcal{J}^E$. Thus the energy current is itself a conserved quantity
implying an infinite energy conductivity at any temperature. Energy
transport in the XXZ chain is purely ballistic.

Spin transport, on the other hand, is a much more complicated
phenomenon. For zero magnetic field, the spin inversion operation
$\mathcal{C}^{-1}
\sigma^z_\ell \mathcal{C}= -\sigma^z_\ell$, $\mathcal{C}^{-1}
\sigma^\pm_\ell \mathcal{C}= \sigma^\mp_\ell$ is a symmetry of the Hamiltonian. It is also
easy to see that $\mathcal{C}^{-1}
\mathcal{J}^E \mathcal{C}= \mathcal{J}^E$. For $h=0$ one can, furthermore, show 
that the transfer matrix $T(\theta)$ in the usual spin $s=1/2$
representation of auxiliary space is itself even under spin inversion
for all spectral parameters $\theta\neq 0,\infty$,
i.e. $\mathcal{C}^{-1} T(\theta)
\mathcal{C}= T(\theta)$. Therefore {\it all} the local conserved charges
$\mathcal{Q}_j$ defined in Eq.~(\ref{Charges}) are even under spin
inversion. The spin current operator, on the other hand, is odd,
$\mathcal{C}^{-1} \mathcal{J}^s \mathcal{C}= -\mathcal{J}^s$. It
follows that for zero magnetic field $\langle \mathcal{J}^s
\mathcal{Q}_j\rangle \equiv 0,\, \forall j$ where $\langle \cdots \rangle$
denotes the thermal average. Therefore the charges $\mathcal{Q}_j$ do
not protect the spin current $\mathcal{J}^s$ from decaying. This leads
to the interesting question whether or not spin transport in the XXZ
chain at zero magnetic field is ballistic or diffusive. We will see in
the following that the answer depends on the anisotropy $\Delta$. For
$\Delta=\cos(\pi/m)$ and $m$ integer, in particular, we will see the
two transport channels coexist. Note that the arguments above do not
apply to the case $h\neq 0$ where spin inversion $\mathcal{C}$ is no
longer a symmetry of the Hamiltonian (\ref{Ham}). In the latter case
it is straightforward to show that a part of the spin current is
protected by the conservation laws (\ref{Charges}) and cannot decay
\cite{ZotosPrelovsek}.

\subsection{Mazur inequality}
To understand more precisely the connection between the Drude weight
and the conserved charges of the considered system, we follow the
approach of Mazur \cite{Mazur} and Suzuki
\cite{Suzuki}. We do so starting with a finite system. This raises some subtle 
questions with regard to the order of taking the limits of system size
and time to infinity. We will get back to this point at the end of
this section.

Let us start by considering the time average of a current-current
correlation in spectral representation
\begin{eqnarray}
\label{Mazur1}
\lim_{\Lambda\to\infty}\frac{1}{\Lambda}\int_0^\Lambda dt\, \langle \mathcal{J}(t)\mathcal{J}(0)\rangle  &=& \sum_{n,m}\frac{\e^{-\beta E_n}}{Z}\langle n|\mathcal{J}|m\rangle\langle m|\mathcal{J}|n\rangle \lim_{\Lambda\to\infty}\frac{1}{\Lambda}\int_0^\Lambda dt\, \e^{it(E_n-E_m)}\nonumber \\ 
&=& \sum_{n,m}^{E_n=E_m} \frac{\e^{-\beta E_n}}{Z}|\langle n|\mathcal{J}|m\rangle|^2 \, ,
\end{eqnarray}
where $Z=\tr\,\{\e^{-\beta H}\}$ is the partition function. Here we have used that taking the limit yields
\begin{equation}
\label{Mazur2}
\lim_{\Lambda\to\infty} \frac{\e^{i\Lambda (E_n-E_m)}-1}{i\Lambda(E_n-E_m)}=\left\{\begin{array}{c} 0,\quad E_n\neq E_m \\ 1,\quad E_n = E_m \end{array} \right. \, .
\end{equation}

Without loss of generality, we can assume that we have a complete set
of Hermitian conserved charges $Q_k$, $[H,Q_k]=0$, which are
orthogonal $\langle Q_k Q_l\rangle =\langle Q_k^2\rangle
\delta_{kl}$. We can then split the current operator into a part which
is diagonal in the energy eigenbasis and a part which is
off-diagonal. The diagonal part can then be expanded in $Q_k$:
\begin{eqnarray}
\label{Mazur3}
&&\mathcal{J} = \sum_k a_k Q_k + \mathcal{J}',\quad \mbox{with}\; \langle n| \mathcal{J}' | m\rangle =0\; \mbox{if}\; E_n=E_m \nonumber \\
&&\Rightarrow \langle Q_l\mathcal{J}\rangle = \sum_k a_k \underbrace{\langle Q_l Q_k\rangle}_{\langle Q_l^2\rangle \delta_{k,l}} + \underbrace{\langle Q_l\mathcal{J}'\rangle}_{=0} \quad \Rightarrow a_l =\frac{\langle Q_l \mathcal{J}\rangle}{\langle Q_l^2\rangle}
\end{eqnarray}
Keeping in mind that $ \langle n| \mathcal{J}' |
m\rangle =0$ if $E_n=E_m$, we can therefore write the time average as
\begin{equation}
\label{Mazur4}
(2.9) = \sum_{n,m}^{E_n=E_m} \frac{\e^{-\beta E_n}}{Z}\sum_{k,l}\frac{\langle \mathcal{J}Q_k\rangle\langle\mathcal{J}Q_l\rangle}{\langle Q_k^2\rangle\langle Q_l^2\rangle}\langle n| Q_k|m\rangle \langle m |Q_l |n\rangle
\end{equation}
The $Q_k$ are diagonal and therefore
\begin{eqnarray}
\label{Mazur5}
\sum_{n,m}^{E_n=E_m} \frac{\e^{-\beta E_n}}{Z}\langle n| Q_k|m\rangle \langle m |Q_l |n\rangle &=& \sum_{n,m} \frac{\e^{-\beta E_n}}{Z}\langle n| Q_k|m\rangle \langle m |Q_l |n\rangle \\
&=& \sum_n \frac{\e^{-\beta E_n}}{Z} \langle n| Q_k Q_l |n\rangle =\langle Q_k Q_l\rangle = \delta_{kl}\langle Q_k^2\rangle \nonumber \, .
\end{eqnarray}
This leads us to the final result
\begin{equation}
\label{Mazur6}
\lim_{\Lambda\to\infty}\frac{1}{\Lambda}\int_0^\Lambda dt\, \langle \mathcal{J}(t)\mathcal{J}(0)\rangle =\sum_k \frac{\langle \mathcal{J}Q_k\rangle^2}{\langle Q_k^2\rangle} \, .
\end{equation}
If we find any conserved charge with $\langle
\mathcal{J}Q_k\rangle\neq 0$ then (\ref{Mazur6}) provides a lower bound for the
time-averaged current-current correlation function in a finite system
because the r.h.s.~of Eq.~(\ref{Mazur6}) is strictly positive.  The
relation is then called the {\it Mazur inequality} and the obtained
bound the {\it Mazur bound}.

In the thermodynamic limit, $N\to\infty$, we expect the
current-current correlation function to equilibrate. If this is the
case, then the time average becomes dominated by the constant
equilibrium value, thus
\begin{equation}
\label{Mazur7}
\lim_{t\to\infty}\lim_{N\to\infty}\frac{1}{2NT}\langle
\mathcal{J}(t)\mathcal{J}(0)\rangle
=\lim_{N\to\infty}\frac{1}{2NT}\sum_k \frac{\langle
\mathcal{J}Q_k\rangle^2}{\langle Q_k^2\rangle} \, .
\end{equation}
In writing Eq.~(\ref{Mazur7}) we take for granted that the Mazur
equality remains valid in the thermodynamic limit, i.e., that we can
take the limit $N\to\infty$ first before taking $t\to\infty$ as is
required in thermodynamics. Physically this is fairly obvious since
the current density-density correlator $\langle
j_\ell(t)j_0(0)\rangle$ is only non-zero (up to exponentially small
tails) within the light cone set by the Lieb-Robinson bounds. I.e.,
for any time $t$ it is sufficient to consider a finite system of size
$N\gg v_{LR} t$ where $v_{LR}$ is the Lieb-Robinson velocity. This
point is discussed in more detail in Ref.~\cite{IlievskiProsen}. We
will see later that Eq.~(\ref{Mazur7}) is proportional to the Drude
weight $D(T)$.

If $\mathcal{J}$ is a local operator---this is the case for the XXZ
chain considered here---then $\langle
\mathcal{J}Q_k\rangle^2\sim N^2$. Therefore only those conserved charges 
contribute to the Mazur bound in the thermodynamic limit for which
\begin{equation}
\label{quasilocal}
\langle Q_k^2\rangle\sim N \, .
\end{equation} 
Operators who fulfill the strict locality condition,
Eq.~(\ref{local}), also fulfill the condition
(\ref{quasilocal}). Additional conserved charges, however, can exist
which are not of the form (\ref{local}) but do fulfill
Eq.~(\ref{quasilocal}). These charges are sometimes called {\it
quasi-local} and play an important role in understanding the spin
transport properties of the XXZ chain. In addition to conserved
charges which are local in the sense of Eq.~(\ref{quasilocal}), every
quantum mechanical system also has an infinite number of non-local
conserved charges. An example are the projectors $P_n=|n\rangle\langle
n|$ onto the extended eigenstates $|n\rangle$ of the system. Such
charges, however, do not affect the transport properties of the
system.

\subsection{Kubo formula}
Next, we want to discuss how to calculate the spin conductivity
$\sigma_s(\omega)$ in linear response and how to relate
Eq.~(\ref{Mazur7}) to the Drude weight. The Kubo formula is obtained
straightforwardly in linear response theory and is given by
\begin{equation}
\label{Kubo1}
\sigma_s(\omega)=\frac{\im}{\omega}\left[\frac{\langle H_{\textrm{kin}}\rangle}{N}-\frac{\im}{N}\int_0^\infty dt\,\e^{\im\omega t} \langle [\mathcal{J}^s(t),\mathcal{J}^s(0)]\rangle\right] \, .
\end{equation}
The first term is the diamagnetic contribution while the second term
is the retarded current-current correlation function. For a derivation
see, for example, the textbook by Mahan \cite{Mahan}. Using again a
spectral representation, we can perform the integral over time and
obtain
\begin{equation}
\label{Kubo2}
\sigma_s(\omega)=\frac{\im}{\omega N}\left[\langle H_{\textrm{kin}}\rangle +\sum_{n,m}\frac{(p_n-p_m)|\langle n|\mathcal{J}^s|m\rangle|^2}{\omega-(E_m-E_n)+\im\delta}\right]
\end{equation}
with $p_n=\exp(-\beta E_n)/Z$ and $\beta=1/T$. We now use the relation
\begin{equation}
\label{Krelation1}
\frac{1}{\omega}\frac{1}{\omega+E}=\frac{1}{E}\left(\frac{1}{\omega}-\frac{1}{\omega+E}\right)
\end{equation}
to split Eq.~(\ref{Kubo2}) into two parts
\begin{equation}
\label{Kubo2_2}
\sigma_s(\omega) = \frac{\im}{\omega N}\left[\langle H_{\textrm{kin}}\rangle +\sum_{n,m}\frac{(p_n-p_m)}{E_n-E_m}|\langle n|\mathcal{J}^s|m\rangle|^2\right] 
-\frac{\im}{N}\sum_{n,m}\frac{(p_n-p_m)}{E_n-E_m}\frac{|\langle n|\mathcal{J}^s|m\rangle|^2}{\omega-(E_m-E_n)} \, .
\end{equation}
The term in the square brackets is the charge or Meissner stiffness
$\Gamma_s$. It can be obtained from the free energy $f(\Phi)$ of an
XXZ ring with a flux $\Phi$ through the ring by $\Gamma_s =
\frac{\partial^2f}{\partial \Phi^2}\big|_{\Phi=0}$. The charge stiffness is
proportional to the superfluid density $n_s(T)$ which is zero in the
thermodynamic limit for a strictly one-dimensional system.

We now take the real part of the last term in Eq.~(\ref{Kubo2_2})
using the relation
\begin{equation}
\label{Cauchy}
\frac{1}{\omega-E} = P\frac{1}{\omega-E} -i\pi\delta(\omega-E) \, 
\end{equation}
to obtain
\begin{eqnarray}
\label{Kubo2_3}
&&\sigma_s'(\omega) = -\frac{\pi}{N}\sum_{n,m}\frac{p_n-p_m}{E_n-E_m}|\langle n|\mathcal{J}^2|m\rangle|^2\delta(\omega-(E_m-E_n)) \\
&=& \frac{\beta\pi}{N}\sum_{E_n=E_m} p_n |\langle n|\mathcal{J}^2|m\rangle|^2 \delta(\omega) + \frac{\pi}{N} \sum_{E_n\neq E_m} \frac{p_n-p_m}{E_m-E_n}|\langle n|\mathcal{J}^2|m\rangle|^2\delta(\omega-(E_m-E_n)) \nonumber \, .
\end{eqnarray}
Comparing with Eq.~(\ref{Drude}) we see that the first term in the
second line is proportional to the Drude weight while the second term
describes the regular part.

Using a spectral representation it is also straightforward to show
that Eq.~(\ref{Kubo2_3}) can be rewritten as a time-dependent
current-current correlation function
\begin{equation}
\label{Kubo2_4}
\sigma_s'(\omega) = \frac{1-\e^{-\beta\omega}}{2\omega N}\int_{-\infty}^\infty \e^{i\omega t}\langle\mathcal{J}^s(t)\mathcal{J}^s(0)\rangle \, .
\end{equation}
This relation is known as the fluctuation-dissipation theorem because
for generic, non-integrable models it connects the current-current
fluctuations to the dissipative part of the conductivity.

For an integrable system, we can split the correlation function into a
ballistic part which persists at infinite times and a regular part
which decays in time
\begin{equation}
\label{Kubo4}
C(t)=\lim_{N\to\infty}\langle \mathcal{J}^s(t)\mathcal{J}^s(0)\rangle/N =\underbrace{\lim_{t\to\infty}\lim_{N\to\infty} \langle \mathcal{J}^s(t)\mathcal{J}^s(0)\rangle/N}_{(\mathcal{J}^s\mathcal{J}^s)_\infty} +C^{\textrm{reg}}_s(t) \, .
\end{equation}
Here $C^{\textrm{reg}}_s(t)$ is a function which vanishes for
$t\to\infty$ and gives a non-singular contribution to the conductivity
$\sigma_s'(\omega)$. Plugging (\ref{Kubo4}) into (\ref{Kubo2_4}) yields
\begin{eqnarray}
\label{Kubo5}
\sigma_s'(\omega)&=& \frac{1-\e^{-\beta\omega}}{2\omega}\int_{-\infty}^\infty dt\,\e^{\im\omega t}\, \left[(\mathcal{J}^s\mathcal{J}^s)_\infty +C^{\textrm{reg}}_s(t)\right] \nonumber \\
&=& 2\pi \frac{(\mathcal{J}\mathcal{J})_\infty}{2T}\delta(\omega) + \frac{1-\e^{-\beta\omega}}{2\omega } C^{\textrm{reg}}_s(\omega) \, .
\end{eqnarray}
Comparing with the definition of the Drude weight and the regular part
of the conductivity (\ref{Drude}) we find the important relation
\begin{equation}
\label{Kubo6}
D_s = \frac{(\mathcal{J}^s\mathcal{J}^s)_\infty}{2T}=\lim_{t\to\infty}\lim_{N\to\infty}\frac{1}{2NT}\langle \mathcal{J}^s(t)\mathcal{J}^s(0)\rangle \, .
\end{equation}
I.e., we have now shown that the expression in (\ref{Mazur7}) is
indeed the Drude weight and that this quantity is directly related to
the part of the current which does not decay. Furthermore,
\begin{equation}
\label{Kubo6_1}
\sigma^{\textrm{reg}}_s(\omega\to 0) = \beta\int_0^\infty dt \, C^{\textrm{reg}}_s(t) = \chi_s(\beta)\mathcal{D}_s
\end{equation}
where we have used the Einstein relation in the second step to
introduce the {\it diffusion constant} $\mathcal{D}_s$ and the static
spin susceptibility $\chi_s$. In addition to the Drude weight which is
related via Eq.~(\ref{Kubo6}) to the part of the current which is
protected by local conservation laws and does not decay in time, there
is thus, in general, also a diffusive part given by the decaying part
of the current with diffusion constant
\begin{equation}
\label{Kubo6_2}
\mathcal{D}_s = \frac{\beta}{\chi(\beta)}\int_0^\infty dt\, \left[C(t)-2T D_s\right] \, .
\end{equation}
We can now combine (\ref{Kubo6}) with the Mazur formula (\ref{Mazur7})
to obtain a bound or the exact Drude weight by considering overlaps of
the conserved charges with the current operator. The advantage of this
approach is that it maps a dynamic onto a static problem. This
approach has been used in
Refs.~\cite{Prosen,ProsenIlievski,PereiraPasquier,ProsenNPB}.

Similar results can also be obtained for the thermal conductivity. A
subtle point is the proper definition of the currents and forces which
cause these currents to flow, see Ref.~\cite{Mahan}. One possible
choice is $\mathcal{J}^s=\frac{M^{11}}{T}\nabla h$ and
$\mathcal{J}^E=M^{22}\nabla\left(\frac{1}{T}\right)$. Comparing with
(\ref{Tcoeffs}) we see that there is an additional factor of $1/T$ in
the definition of the thermal conductivity $\kappa_{\textrm{th}}$. For
the thermal Drude weight at zero field one finds, in particular,
\begin{equation}
\label{Kubo7}
D_{\textrm{th}} = \lim_{t\to\infty}\lim_{N\to\infty}\frac{1}{2NT^2}\lim_{t\to\infty}\langle \mathcal{J}^E(t)\mathcal{J}^E(0)\rangle =\lim_{N\to\infty}\frac{\langle (\mathcal{J}^E)^2\rangle}{2NT^2} \, 
\end{equation}
where we have used in the last step that $[\mathcal{J}^E,H]=0$ for the
XXZ chain.

\section{Thermal Drude weight}
The thermal Drude weight is particularly easy to calculate because it
is given by the static expectation value of a conserved charge, see
(\ref{Kubo7}). In the following we briefly sketch how to obtain
$D_{\textrm{th}}$ using the standard thermodynamic Bethe ansatz (TBA)
formalism for anisotropies $\Delta=\cos(\gamma)$ with
$\gamma=\pi/m$. We note that the first derivation of the thermal Drude
weight was carried out by Kl\"umper and Sakai \cite{KluemperSakai}
using the quantum transfer matrix formalism. The latter approach has
the advantage that the string hypothesis is not needed and results for
arbitrary $\Delta$ are obtained.

We consider only the case $h=0$. First, we define a generalized
partition function and generalized free energy 
\begin{equation}
\label{JE1}
Z=\tr \exp(-\beta H + \lambda J^E), \quad f(\beta,\lambda)=-\frac{T}{N}\ln Z \, .
\end{equation}

In TBA we can write this free energy density as 
\begin{equation}
\label{FreeE}
f(\beta,\lambda) =-\frac{T}{2\pi}\sum_{\ell=1}^m\int d\theta\, \varepsilon_{\ell}(\theta)\sigma_\ell\ln[1+\eta_\ell^{-1}(\theta)] .
\end{equation}
Here $\varepsilon_\ell$ are the bare eigenenergies. The variables
$\sigma_\ell = \text{sign}(g_\ell)$ are the signs of auxiliary
rational numbers associated to string solutions as defined
in~\cite{tak99}. For the case of anisotropy $\gamma=\pi/m$ the
$g_\ell$ have a particularly simple relation to string length $n_\ell$
\begin{equation}
\label{sgnFct}
g_\ell = m - n_\ell ,\, n_\ell = \ell \, \text{ for } \, \ell= 1, \dots, m-1 \text{ and } \, g_m =-1,\, n_m=1.
\end{equation}
The functions $\eta_\ell=\rho^h_\ell/\rho_\ell$ are defined by the ratio
of hole density $\rho^h_\ell$ and particle density $\rho_\ell$ of the
$\ell$-th particle (string) and fulfill the coupled TBA equations
\begin{align}
\ln\eta_\ell(\theta) &=\beta \varepsilon_\ell +\lambda j_\ell^E + \sum_\kappa \int d\mu K_{\ell \kappa}(\theta-\mu) \sigma_\kappa  \ln(1+\eta_\kappa^{-1}(\mu)),\nonumber\\
\label{basic}
&\equiv \beta \varepsilon_\ell +\lambda j_\ell^E + \left[K * \sigma \ln(1+\eta^{-1}) \right]_\ell
\end{align}
with an integration kernel $K$, and '$*$' denoting a convolution and
sum over Bethe strings. Here $j_\ell^E= \partial_\theta
\varepsilon_\ell = \partial^2_\theta p_\ell=p''_\ell$ where
$p(\theta)$ is the momentum. To express the results in a more compact form, it is useful to define the following dressed quantities
\begin{equation}
\label{dressIntFn}
\dr{\varepsilon}_\ell = \varepsilon_\ell - \left[ K * \sigma \vartheta \dr{\varepsilon}\right]_\ell,  \quad \dr{j}^E_\ell = j^E_\ell - \left[ K * \sigma \vartheta \dr{j}^E\right]_\ell 
\end{equation}
where we have defined the Fermi factor
$\vartheta_\ell=1/(1+\eta_\ell)=\rho_\ell/(\rho_\ell +
\rho^h_\ell)$. It is also useful to realize the following simple
relation of the dressed quantities to the logarithmic derivatives
of the $\eta$-functions
\begin{equation}
\label{eta}
\partial_{\beta} \log \eta_\ell(\theta) = \dr{\varepsilon}_\ell(\theta), \quad \partial_{\lambda} \log \eta_\ell(\theta) = \dr{j}^E_\ell(\theta) \, . 
\end{equation}

It is now straightforward to obtain the expectation value needed to calculate the thermal Drude weight
\begin{eqnarray}
\label{QQ}
&& \langle (J^E)^2\rangle /N = -\frac{1}{T}\partial_{\lambda}^2 f(\beta,\lambda)|_{\lambda=0} =\frac{1}{2\pi}\sum_\ell\int d\theta\, \varepsilon_\ell\sigma_\ell\partial_{\lambda}^2\ln(1+\eta_\ell^{-1}) \nonumber \\
&& = \frac{1}{2\pi}\sum_\ell\int d\theta\,\sigma_\ell\vartheta_\ell(1-\vartheta_\ell)\dr{\varepsilon}_\ell (\dr{j}^E_\ell)^2 = \sum_\ell \int d\theta\, \rho_\ell (1-\vartheta_\ell) (\dr{j}^E_\ell)^2 \, .
\end{eqnarray}
Here we have used several identities which are described in Refs.~\cite{Zotos2017,UrichukOez}. 

\begin{figure}
\includegraphics*[width=0.99\textwidth]{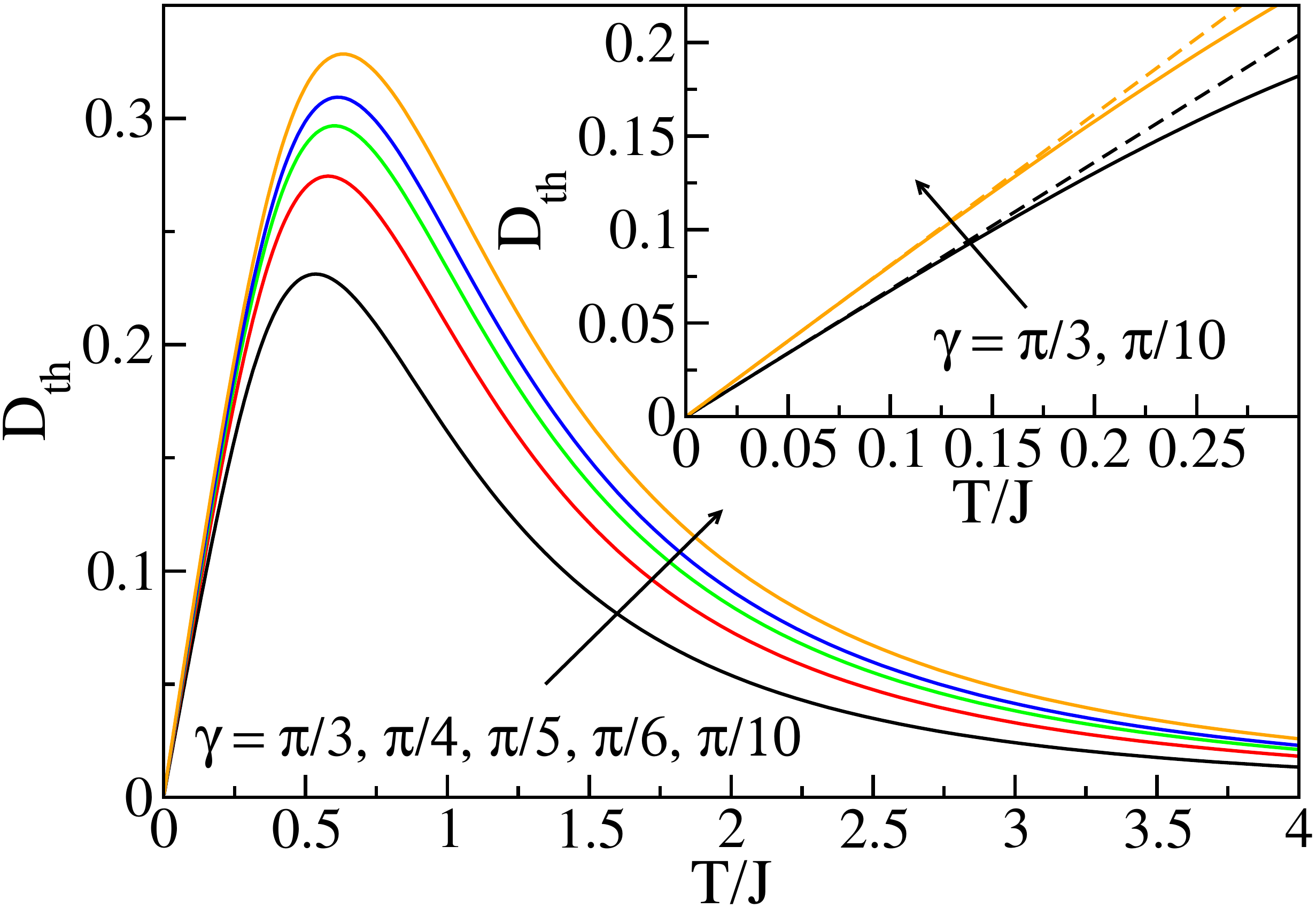}
\caption{\label{Fig1} $D_{\textrm{th}}(T)$ for different anisotropies. The inset compares the full result to the low-temperature asymptotics (dashed lines), see Eq.~(\ref{kappa}).}
\end{figure}

In Fig.~\ref{Fig1} we show results for the thermal Drude weight
$D_{\textrm{th}}=\langle\left(\mathcal{J}^E\right)^2\rangle/2NT^2$
for anisotropies $\Delta=\cos(\pi/m)$ as a function of temperature.
At low temperatures one finds that both the thermal Drude weight
$D_{\textrm{th}}(T)$ and the specific heat $C(T)$ scale linearly with
temperature
\begin{equation}
\label{kappa}
D_{\textrm{th}} =\frac{\pi v}{6} T,\quad C=\frac{\pi}{3v}T,\quad \frac{D_{\textrm{th}}}{C}=\frac{v^2}{2} \, ,
\end{equation}
where $v=J\pi\sin\gamma/2\gamma$ is the velocity of the elementary
excitations.

We can now ask if unusual heat transport properties can be observed in
experiments on spin-$1/2$ chain compounds in which integrability will
be broken by lattice vibrations, impurities, and interchain
couplings. Before discussing this point further, two important
comments are in order. If one measures the thermal conductivity at
finite magnetic field, then the thermal current consists of energy {\it
and} spin current contributions: $\mathcal{J}^{\textrm{th}} =
\mathcal{J}^E - h\mathcal{J}^s$. Experimental measurements of the heat
conductivity are often done in a setup where the spin current
vanishes, $\mathcal{J}^s=0$. In this case the heat conductivity is
redefined
\begin{equation}
\label{magneto}
K=\kappa_{\textrm{th}} -\frac{1}{T}\frac{\left(C^{\textrm{th}}_s\right)^2}{\sigma_s}
\end{equation}
where the second term is called the magnetothermal correction
\cite{PsaroudakiZotos}. 
Eq.~(\ref{magneto}) is based on the assumption that the relaxation
times for energy and spin transport are the same which might not
necessarily be true for a real material. Leaving such issues aside,
one might expect that in a system which is close to an integrable one,
heat currents are decaying slowly and mean free paths are long. This
is indeed what seems to have been seen in a number of experiments
\cite{SologubenkoGianno,SologubenkoBerggold,KohamaSologubenko}. For the copper-oxide spin chain 
compounds Sr$_2$CuO$_3$ and SrCuO$_2$, for example, it has been
observed that the heat conductivity along the chain direction is about
an order of magnitude larger than in the perpendicular directions. A
natural explanation appears to be that there is heat transport due to
phonons in all directions while only in the chain direction there is
an additional contribution due to magnetic excitations which decays
only very slowly. Obtaining a detailed understanding of the heat
transport as measured experimentally is, however, a complex and still
somewhat open issue. It requires an identification of the dominant
relaxation processes and a formalism to incorporate such scattering
mechanisms in the calculation of the thermal conductivity.

\section{The Spin Conductivity}
In this last part, I want to discuss the spin conductivity of the XXZ
chain at zero magnetic field. As already discussed, in this case none
of the conserved charges (\ref{Charges}) derived from the regular
transfer matrix has any overlap with the spin current because of the
spin-flip symmetry $\mathcal{C}$. The Mazur inequality (\ref{Mazur7})
therefore apparently does not provide a non-zero bound. This raises
the question whether or not spin transport in the integrable XXZ chain
has a ballistic component. Various different approaches have been used
so far to try to directly compute the Drude weight: (1) Starting from
the spectral representation of the Kubo formula (\ref{Kubo2}) and
comparing this with the change of the eigenenergies $E_n$ of the
Hamiltonian (\ref{Ham}) when threading a static magnetic flux $\Phi$
through an XXZ ring one finds
\begin{equation}
\label{Kohn}
D=\frac{1}{2NZ}\sum_n \e^{-E_n/T}\frac{\partial^2E_n(\Phi)}{\partial \Phi^2}\bigg|_{\Phi=0}
\end{equation}
with $Z$ being the partition function. This is a generalization of the
Kohn formula \cite{Kohn} to finite temperatures \cite{CastellaZotos}. 
For zero temperature, in particular, the Drude weight can be 
obtained simply from the ground state energy of the system with an
added flux \cite{ShastrySutherland} leading to
\begin{equation}
\label{Dzero}
D(T=0) = \frac{\pi\sin\gamma}{8\gamma(\pi-\gamma)}.
\end{equation}
For finite temperatures, the formula (\ref{Kohn}) has been used in
Ref.~\cite{Zotos} to calculate $D(T)$ for anisotropies $\gamma=\pi/m$
on the basis of the thermodynamic Bethe ansatz (TBA). The high- and
low-temperature limits have then been analyzed in
Ref.~\cite{KluemperJPSJ}. (2) A completely different approach is based
on constructing a set of quasi-local charges---different from the ones
in Eq.~(\ref{Charges})---that have finite overlap with the current
operator and to evaluate the r.h.s.~of Eq.~(\ref{Mazur7}), see for
example
Refs.~\cite{Prosen,ProsenIlievski,PereiraPasquier,ProsenNPB}. A major
difficulty in this approach is the evaluation of the correlators at
finite temperatures. So far, only the high-temperature limit has been
analyzed analytically \cite{ProsenIlievski} resulting in
\begin{equation}
\label{Dinf}
\lim_{T\to\infty} 16TD=J^2\frac{\sin^2(\pi n/m)}{\sin^2(\pi/m)}\left(1-\frac{m}{2\pi}\sin(2\pi/m)\right).
\end{equation}
Here the equal sign is only correct if the set of conserved charges
used is complete which is a point which is difficult to prove. It has,
however, been shown that the above result agrees with the
high-temperature limit of the TBA result obtained using the Kohn
formula \cite{UrichukOez} which might give us some confidence that
(\ref{Dinf}) is not just a lower bound but indeed exhaustive. Note
that the Drude weight in the high-temperature limit has a fractal
character according to Eq.~(\ref{Dinf}), while $D(T=0)$ depends
smoothly on anisotropy, see Eq.~(\ref{Dzero}). This is opposite to our
usual expectations that thermal fluctuations lead to a smoothening of
the expectation values of observables as function of some parameter of
the model. (3) A third approach has recently been proposed based on a
generalized hydrodynamics (GHD) formulation where the continuity
equations
\begin{equation}
\label{Cont}
\partial_t \langle Q_n\rangle +\partial_x \langle J_n\rangle =0
\end{equation}
lead to the so-called Bethe-Boltzmann equations
\cite{BertiniCollura,CastroAlvaredoDoyon,BulchandaniVasseur,DoyonSpohn}
\begin{equation}
\label{BB}
\partial_t\rho_{\xi,\ell}(\theta)+\partial_x\left(v_{\xi,\ell}(\theta)\rho_{\xi,\ell}(\theta)\right)=0 \, .
\end{equation}
Here the current $J_n$ is being related to the velocity $v_{\xi,\ell}$
and density $\rho_{\xi,\ell}$ of quasi-particle excitations. $\xi=x/t$
describes a set of rays along which a local equilibration is assumed
to occur. The advantage of this formulation is that also dynamics far
from equilibrium can be investigated. (4) Very recently, a first
principle calculation of the Drude weight starting directly from the
operator expression of the spin current has been presented
\cite{UrichukOez}. Here the only assumption remaining is related to
the existence of a complete set of conserved charges, similar to the
assumption used in the derivation of the Mazur inequality.

Since GHD has already been discussed at this Les Houches summer
school, I will spend the last part of this lecture series on
introducing an effective low-energy approach. In contrast to Bethe
ansatz methods, this will allow to obtain a physical picture of the
spin conductivity not only in integrable but also in generic
spin-chain models. Furthermore, for the integrable XXZ chain we will
be able to directly connect the ballistic and diffusive transport
channels to each other.

\subsection{Bosonization}
Let me very briefly recapitulate the idea of bosonization. We start
from the fermionic Hamiltonian (\ref{Ham2}) and take the continuum limit 
\begin{equation}
\label{Bos1}
c_j\to \Psi(x) = \e^{ik_F x} \Psi_R(x) + \e^{-ik_F x}\Psi_L(x), \quad
\Psi_{R,L}(x)=\frac{1}{\sqrt{N}}\sum_{k=-\Lambda}^\Lambda
c_{kR,L}\e^{\pm ikx}
\end{equation}
where $\Psi_{R,L}$ are the right and left movers obtained by
linearizing the dispersion around the Fermi points and $\Lambda$ is a
momentum cutoff. The important point is that particle-hole excitations
with momentum $q$ now all have the same energy, e.g., $E_R(q)=v(k+q)-vk
= vq$ is independent of $k$ with $v$ being the velocity. Collective
excitations of particle-hole type can therefore be represented by a
bosonic operator, $\sum_k c^\dagger_{k+q}c_k\sim b_q$, and the
interacting Hamiltonian (\ref{Ham2}), which is quartic in the fermionic
operators, becomes a {\it quadratic bosonic theory} at low
energies. The correction terms to the quadratic theory are all
irrelevant in a renormalization group sense in the critical regime
$-1<\Delta <1$. For the purpose of calculating the conductivity it is
convenient to use bosonic fields which are related to the right and
left movers by
\begin{equation}
\label{Bos3}
\Psi_{R,L}\propto \frac{1}{\sqrt{2\pi\alpha}}\e^{-\im\sqrt{2\pi}\varphi_{R,L}},\quad \varphi_{R,L}=\frac{1}{\sqrt{2}}(\tilde\theta\mp\tilde\phi) \, ,
\end{equation}
where $\alpha\sim k_F^{-1}$ is a short-distance cutoff and we have
introduced canonically conjugated fields
$[\tilde\phi(x),\partial_{x'}\tilde\theta(x')]=\im\delta(x-x')$. The
interaction now merely leads to a rescaling of these fields, $\tilde
\phi = \sqrt{K/2}\phi$ and $\tilde\theta = \sqrt{2/K}\theta$, leading
to a Hamiltonian
\begin{equation}
\label{Bos2}
H=\frac{v}{2}\int dx\, \left[(\partial_x\phi)^2 +(\partial_x\theta)^2\right] +\lambda \int dx\, \cos(\sqrt{8\pi K}\phi) \, .
\end{equation}
The first term describes the free theory while the second term with
scaling dimension $2K$ represents irrelevant Umklapp scattering. The
Luttinger parameter $K$ and the velocity $v$ can be determined for the
integrable XXZ chain by calculating static properties such as the
specific heat and the susceptibility using the field theory
(\ref{Bos2}) and the Bethe ansatz and comparing the results. This
leads to
\begin{equation}
\label{Bos4}
v=\frac{J\pi}{2}\frac{\sqrt{1-\Delta^2}}{\arccos\Delta}=\frac{J\pi}{2}\frac{\sin\gamma}{\gamma},\quad
K=\frac{\pi}{\pi-\arccos\Delta}=\frac{\pi}{\pi-\gamma} \, .
\end{equation}
Note that in this notation $K=2$ at the free Fermi point $\Delta=0$,
and $K=1$ at the isotropic point $\Delta=1$. 

The spin current density is given by
$j^s=J(\Psi^\dagger_L\Psi_L-\Psi^\dagger_R\Psi_R)$ in terms of the
left and right movers. Since the free bosonic Hamiltonian conserves
the right and left particle densities separately, the spin current
will not relax. It is thus important to also take the last term in
Eq.~(\ref{Bos2}) into account. It describes Umklapp scattering
\begin{equation}
\label{Bos5}
\sim \e^{-i2k_F(2x+1)}\Psi_R^\dagger(x)\Psi_L(x)\Psi_R^\dagger(x+1)\Psi_L(x+1)+h.c. \, 
\end{equation}
where two left movers scatter to two right movers and vice versa. In
general, this term oscillates $\sim\exp(i4k_Fx)$ but is
non-oscillating at half-filling (zero magnetic field) where
$k_F=\pi/2$. While this term is formally irrelevant for $-1<\Delta<1$
it can relax the current and therefore has to be treated with care.

\subsection{Results}
We now want to evaluate the Kubo formula (\ref{Kubo1}). We can couple
the fermions to the electromagnetic potential $A$ by a Peierls
substitution $\Pi=\partial_x\theta \to \Pi-\sqrt{K/2\pi}A$. One then
finds
\begin{eqnarray}
\label{Cond3}
\frac{\partial H}{\partial A}\bigg|_{A=0} &=& \int dx\, j^s(x)\quad \mbox{with}\quad j^s=-v\sqrt{\frac{K}{2\pi}}\Pi=-\sqrt{\frac{K}{2\pi}}\partial_t\phi\, , \\
\frac{\partial^2 H}{\partial A^2}\bigg|_{A=0} &=& \langle H_{\textrm{kin}}\rangle = \frac{vK}{2\pi}L \nonumber \, 
\end{eqnarray}
using $\partial_x\theta = v^{-1}\partial_t\phi$. Here $L=Na$ with $a$
being the lattice constant. The second line is the diamagnetic
term. The Kubo formula then reads
\begin{equation}
\label{Cond4}
\sigma_s(q,\omega)=\frac{\im}{\omega}\left[\frac{vK}{2\pi}+\langle \mathcal{J}^s \mathcal{J}^s\rangle^{\textrm{ret}}(q,\omega)\right] \, .
\end{equation}
By partial integration and using the canonical commutation relations
one finds
\begin{equation}
\label{Cond5}
\langle \partial_t\phi\partial_t\phi\rangle^{\textrm{ret}}(q,\omega)=-v+\omega^2\langle \phi\phi\rangle^{\textrm{ret}}(q,\omega) \, .
\end{equation}
Putting this into (\ref{Cond4}) we see that the diamagnetic term is
cancelled and we are left with the following simple Kubo formula for
the spin conductivity within the bosonized theory
\begin{equation}
\label{Cond6}
\sigma_s(q,\omega)=\frac{vK}{2\pi}\im\omega\langle \phi\phi\rangle^{\textrm{ret}}(q,\omega) \, .
\end{equation}
The only quantity required to obtain the conductivity is thus the
retarded correlation function of the basic bosonic field. For the free
bosonic model without the Umklapp term ($\lambda=0$ in
Eq.~(\ref{Bos2})), we just find the standard free boson propagator
\begin{equation}
\label{Cond7}
\langle\phi\phi\rangle^{\textrm{ret}}(q,\omega)=\frac{v}{\omega^2-v^2q^2} \, 
\end{equation}
leading to a Drude weight
\begin{equation}
\label{Cond8}
D\delta(\omega)=\frac{1}{2\pi}\lim_{\omega\to 0}\lim_{q\to 0}\sigma'(q,\omega)=\frac{Kv}{4\pi^2}\textrm{Re}\left(\frac{\im}{\omega+\im\epsilon}\right)=\frac{Kv}{4\pi}\delta(\omega) \, 
\end{equation}
which does agree with the BA result (\ref{Dzero}). Note that at this
level of approximation there is no regular part of the conductivity
and no temperature dependence of the Drude weight. Taking into account
band curvature terms will introduce a temperature dependence of the
Drude weight but only the Umklapp term can lead to a relaxation of the
current. For the conductivity this operator is dangerously irrelevant
and will completely change the transport properties of the theory. To
see this it is sufficient to calculate the propagator to second order
in perturbation theory in the Umklapp scattering
\begin{equation}
\label{Cond9}
\langle\phi\phi\rangle^{\textrm{ret}}(q,\omega)=\frac{v}{\omega^2-v^2q^2-\Pi^{\textrm{ret}}(q,\omega)} \, 
\end{equation}
where $\Pi^{\textrm{ret}}(q,\omega)$ is the self energy. This is a
standard calculation and we just present the result here
\begin{equation}
\label{Cond10}
\sigma(q,\omega)=\frac{vK}{2\pi}\frac{\im\omega}{\omega^2-v^2q^2+2\im\varGamma\omega} \, .
\end{equation}
Here $\varGamma\sim \lambda^2 T^{4K-3}$ is a relaxation rate which
vanishes for $T\to 0$. For the integrable XXZ model, $\varGamma$ can
be determined exactly \cite{Lukyanov} and there are therefore no free
parameters in (\ref{Cond10}) in this case. Here we just want to
understand the physics qualitatively. Considering, in particular, the
real part of the conductivity at $q=0$ we find
\begin{equation}
\label{Cond11}
\sigma'(\omega)=\frac{vK}{2\pi}\frac{2\varGamma}{\omega^2+(2\varGamma)^2} \, .
\end{equation}
The Drude weight broadens to a Lorentzian with width $\sim T^{4K-3}$
at any finite temperature. While this is in fact the expected behavior
for a generic non-integrable model, we are now missing the
finite-temperature Drude weight which we know does exist in the
integrable XXZ chain because of the quasi-local charges which protect
a part of the spin current from decaying.

This should not come as a surprise: In the derivation of the
low-energy effective theory, the existence of an infinite set of
(quasi-)local conserved charges $Q_n$ has not been taken into
account. The requirement $[H,Q_n]=0$ corresponds, in general, to a
fine-tuning of the bosonic Hamiltonian. As has been shown in
Ref.~\cite{PereiraSirkerJSTAT} this can, for example, lead to the
absence of certain irrelevant terms which are kinematically allowed and
therefore expected to be present in a generic model. A full
understanding of the structure of the low-energy Hamiltonian for the
integrable XXZ chain is, however, still lacking. Here we will instead
use a different approach. If there is a conserved charge with finite
overlap with the current, then we can separate this current into two
parts
\begin{equation}
\label{Cond12}
\mathcal{J}^s=\underbrace{\frac{\langle\mathcal{J}^sQ\rangle}{\langle Q^2\rangle}Q}_{\mathcal{J}^s_\parallel} + \mathcal{J}^s_\perp \, .
\end{equation}
Then $\mathcal{J}^s_\perp$ will decay due to Umklapp scattering while
$\mathcal{J}^s_\parallel$ is protected. More formally, this approach can
be implemented using a memory matrix approach, see
Ref.~\cite{SirkerPereira,SirkerPereira2}. The conductivity then becomes
\begin{equation}
\label{Cond13}
\sigma'_s(\omega) = \underbrace{\frac{vK}{2}\frac{y}{1+y}}_{2\pi D_s(T)}\delta(\omega) + \underbrace{\frac{vK}{\pi}\frac{\varGamma}{\omega^2+4(1+y)^2\varGamma^2}}_{\sigma'_{\textrm{reg}}(\omega)}
\end{equation}
with
\begin{equation}
\label{Cond14}
\frac{y}{1+y} = \frac{\langle \mathcal{J}^sQ\rangle^2}{\langle (\mathcal{J}^s)^2\rangle\langle Q^2\rangle} \, 
\end{equation}
and $\langle (\mathcal{J}^s)^2\rangle/LT = vK/2\pi$. Note that the
Drude weight $D_s$ obtained from Eqs.~(\ref{Cond13}) and
(\ref{Cond14}) is consistent with the Mazur equation
(\ref{Mazur7}). Note, furthermore, that for $y\to\infty$ and thus
$y/(1+y)\to 1$ we recover the Drude weight $D_s=vK/(4\pi)$ which
therefore corresponds to the case of a fully conserved current. For
$y$ finite, on the other hand, Eq.~(\ref{Cond13}) describes a {\it
coexistence of ballistic and diffusive transport}. Finally, we can
also check that (\ref{Cond13}) fulfills the f-sum rule $\int d\omega\,
\sigma'_s(\omega)=vK/2$.

Conversely, we can also use (\ref{Cond13}) to express $y$ by the Drude
weight $D_s$ leading to
\begin{equation}
\label{eq3}
y = \frac{4\pi D_s(T)}{vK-4\pi D_s(T)}, \quad 1+y = \frac{vK}{vK-4\pi D_s(T)} 
\end{equation}
with $D_s(0)=vK/(4\pi)$. The regular part of the conductivity at frequency zero then reads
\begin{equation}
\label{eq4}
\sigma'_{\textrm{reg}}(\omega=0) = \frac{vK}{4\pi}\frac{1}{(1+y)^2\varGamma}=\frac{(vK-4\pi D(T))^2}{4\pi vK \varGamma} \, .
\end{equation}
For $\gamma=\pi/m$, the TBA calculations in
Refs.~\cite{Zotos,UrichukOez} have shown that at low temperatures the
Drude weight behaves as $D_s(T)=D_s(0)-\alpha T^{2K-2}$ where $\alpha$
depends on the anisotropy $\gamma$. Furthermore, the relaxation rate
due to Umklapp scattering can be expressed as $\varGamma=\varGamma_0
T^{4K-3}$ where $\varGamma_0$ is a function of anisotropy and is known
exactly, see Ref.~\cite{SirkerPereira,SirkerPereira2}. The regular part of the
conductivity at low temperatures is therefore given by
\begin{equation}
\label{eq5}
\sigma'_{\textrm{reg}}(\omega=0) = \frac{4\pi\alpha^2}{vK \varGamma_0}\frac{1}{T} \, .
\end{equation}
We can now use the Einstein relation to define the diffusion constant 
\begin{equation}
\label{eq6}
\mathcal{D}_s\equiv\frac{\sigma'_{\textrm{reg}}(\omega=0)}{\chi_s} = \frac{8\pi^2\alpha^2}{K^2 \varGamma_0}\frac{1}{T} \, ,
\end{equation}
where $\chi_s(T)$ is the spin susceptibility and we have used the
low-temperature result $\chi_s=K/2\pi v$. The diffusion constant thus
diverges as $1/T$ for $T\to 0$. Note that this derivation uses the
Bethe ansatz result for anisotropies $\Delta=\cos(\pi/m)$ and is thus
only valid for these discrete anisotropies. Furthermore, the
relaxation rate $\varGamma=\varGamma_0 T^{4K-3}$ has only been
calculated to second order in Umklapp scattering so Eq.~(\ref{eq6}) is
only expected to be an upper bound for the exact diffusion constant at
low temperatures. A formula to calculate the exact diffusion constant
at anisotropies $\Delta=\cos(\pi n/m)$ has recently been conjectured
in Ref.~\cite{deNardisBernardDoyon} based on an extension of
GHD. Numerically, these predictions can be tested by calculating the
diffusion constant directly from the current-current correlation
function, see Eq.~(\ref{Kubo6_2}). In such numerical calculations, the
main problem is to reach sufficiently long times to obtain reliable
results for the integral over the time-dependent current-current
correlation function. This problem is particularly severe at low
temperatures where the current-current correlation function decays
very slowly towards its long-time value $\lim_{t\to\infty}\langle
\mathcal{J}^s(t)\mathcal{J}^s(0)\rangle = 2NT D_s(T)$.

\begin{figure}
\begin{center}
\includegraphics[width=0.7\textwidth]{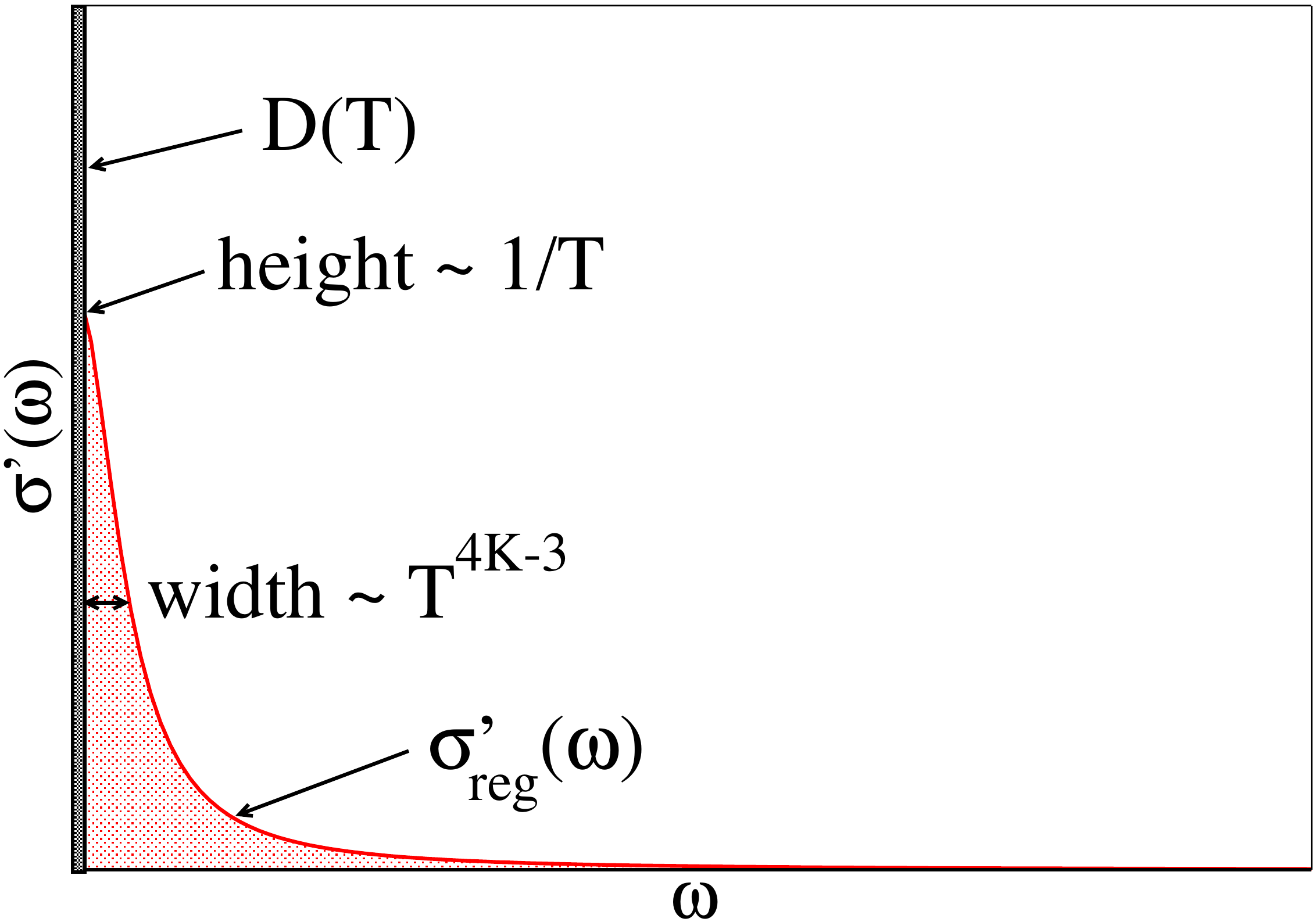}
\end{center}
\caption{\label{Fig_final} At finite temperatures and anisotropies $\Delta=\cos(\pi/m)$ there is a coexistence of ballistic and diffusive transport in the XXZ chain: The Drude peak sits on top of a narrow Lorentzian with width $\sim T^{4K-3}$.}
\end{figure}

Overall, we have obtained the following picture for the spin
conductivity $\sigma'_s(\omega)$ of the XXZ chain at $h=0$ and small
frequencies $\omega$: At $T=0$ there is only a Drude peak
$D=vK/(4\pi)$ and no regular part because Umklapp scattering is
inactive. At $T>0$, on the other hand, we have a coexistence of
ballistic and diffusive transport. This coexistence manifests itself
in a Drude peak on top of a narrow Lorentzian with width $\sim
T^{4K-3}$ and height $1/T$. The weight of the Lorentzian is therefore
$\sim T^{4K-4}$ and vanishes for $T\to 0$ if
$0<\Delta=\cos(\pi/m)<1$. This situation is shown pictorially in
Fig.~\ref{Fig_final}.

\section{Conclusion}
To summarize, I have introduced the basic framework to calculate
transport in the linear response regime. For integrable models,
transport can be unusual in the sense that the current itself or part
of the current is protected by a conservation law leading to an
infinite dc conductivity even at finite temperatures. It is important
to stress that the ideal conductivity in this case is not related
to superconductivity: the superfluid density is zero and there is no
Meisner effect.

For the integrable XXZ spin chain in the critical regime, concrete
results for the thermal and the spin conductivity at anisotropies
$\gamma=\pi/m$ have been derived. These results can be easily
generalized to $\gamma=n\pi/m$ with $n,m$ coprime and integer. Note
that while the TBA-type approaches used here rely on having finite string
lengths and can therefore not be applied if $\gamma/\pi$ is irrational,
we can approximate any irrational number by a rational one to
arbitrary precision. The result for the infinite temperature spin
Drude weight (\ref{Dinf}), for example, does have a well-defined limit
$16TD_s=2\sin^2(\gamma)/3$ for $\gamma$ irrational. This suggests that
the XXZ chain does show an infinite dc conductivity for all
anisotropies $-1<\Delta<1$ and all temperatures.

Left out of these lectures has been the gapped regime of the XXZ
chain, $|\Delta|> 1$, and the isotropic antiferromagnet,
$\Delta=1$. For the thermal Drude weight nothing changes
qualitatively because $\mathcal{J}^E$ itself is conserved. The
quasi-local charges which protect part of the spin current, on the
other hand, become non-local for $|\Delta|>1$ and the spin transport
becomes diffusive
\cite{IlievskideNardisMedenjak,deNardisBernardDoyon}. Right at the isotropic point, $\Delta=1$, 
numerical calculations point to super-diffusive transport with a
dynamical critical exponent $z=2/3$ \cite{LjubotinaZnidaric}. While a
mostly coherent picture of spin transport in the XXZ chain has started
to emerge in the last ten years based on a number of different
analytical and numerical methods, these very recent results for the
isotropic point show that this picture is not quite complete yet and
that this topic deserves further study.

\section*{Acknowledgements}
I am grateful to my colleagues and co-authors on a number of related
research articles for sharing their insights into this subject. In
particular, I would like to thank I. Affleck, R.G. Pereira, and
A. Kl\"umper. I also thank A. Urichuk for providing the data
for the thermal Drude weight.

I acknowledge support by the Natural Sciences and Engineering Research
Council (NSERC) through the Discovery Grants program and by the German
Research Foundation (DFG) via the Research Unit FOR 2316.


\nolinenumbers

\end{document}